# NANO LETTERS

Cite This: *Nano Lett.* 2019, 19, 8855−8861

pubs.acs.org/NanoLett

Letter# Deterministic Deposition of Nanoparticles with Sub-10 nm Resolution

Stefan Fringes, C. Schwemmer, Colin D. Rawlings, and Armin W. Knoll*

IBM Research − Zurich, Säumerstrasse 4, 8803 Rüschlikon, Switzerland**S** Supporting Information

**ABSTRACT:** Accurate deposition of nanoparticles at defined positions on a substrate is still a challenging task, because it requires simultaneously stable long-range transport and attraction to the target site and precise short-range orientation and deposition. Here we present a method based on geometry-induced energy landscapes in a nanofluidic slit for particle manipulation: Brownian motors or electro-osmotic flows are used for particle delivery to the target area. At the target site, electrostatic trapping localizes and orients the particles. Finally, reducing the gap distance of the slit leads 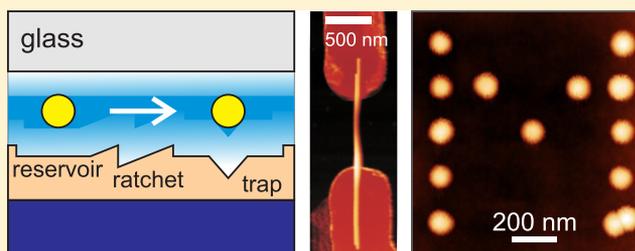 sequentially to a focusing of the particle position and a jump into adhesive contact by several nanometers. For 60 nm gold spheres, we obtain a placement accuracy of 8 nm. The versatility of the method is demonstrated further by a stacked assembly of nanorods and the directed deposition of InAs nanowires.

**KEYWORDS:** *Nanofluidics, nanofabrication, nanowires, nanoparticle assembly*Nanoparticles and nanowires are readily available through bottom-up synthesis in a variety of shapes and with exceptional properties often surpassing the performance of top-down fabricated structures. Their precise deposition is required for the fabrication of novel (quantum-) nanoelectronic circuits[1−3] and is important for applications in bioelectronics.[4] However, a high-density directed-assembly method is still not available for high aspect ratio objects. The task is far from trivial, since it involves several steps: the particles have to be transported to the location of interest, they need to be aligned and immobilized at the correct location, and nonspecific adsorption has to be prevented.

A variety of assembly methods exist,[5−7] with two prominent examples being capillary assembly[5] and dielectrophoretic deposition.[6] In capillary assembly, the receding meniscus of a nanoparticle suspension pushes the particles into topographical traps. The final orientation depends on the drying process, and complex template fabrication is required for precise deposition.[8] Moreover, the high density at the meniscus leads to ordering of asymmetric particles and correct orientation could only be demonstrated for particles with small aspect ratios.[8−10] In dielectrophoretic trapping and deposition, a high frequency electric field between predefined electrodes is used to create an attractive trapping and deposition force. The self-limiting nature of the process allows for the deposition of single high aspect ratio objects.[6] Moreover, it can be combined with capillary assembly to facilitate transport to the trapping sites and the drying process.[11] However, the achievable complexity is limited by the area occupied by the powered electrodes.

Geometry-induced electrostatic trapping (GIT) is capable of confining nanoscale gold particles[12,13] with a diameter down to 10 nm[14] and even single 60 bp DNA molecules and proteins.[15] This trapping mechanism exploits the repulsive force of like charged particles confined between a flat and a patterned surface. The geometry was defined either by a recess topography in silicon oxide[12,13] or by a nanopipette[14] in close proximity. By adapting the trap shape, the orientation of trapped nanorods was controlled and could be switched by external fields[16] for data storage devices.[17] Recently, we used nanometer precise grayscale patterns written by thermal probe lithography for GIT pattern definition in polyphthalaldehyde (PPA). The thus defined energy landscapes enable fast and size selective transport of gold nanoparticles[18] in rocked Brownian motors. Moreover, a current reversal[19] was observed in this overdamped system.

Here, we propose to use GIT for the alignment and precise deposition of nanoparticles at reference positions on a substrate. We first describe the parallel trapping and depositon of 60 nm Au spheres at predefined positions. We observe the details of the trapping and deposition steps with high temporal and spatial resolution using iSCAT imaging.[20] At sufficiently small separations, we observe how the particles jump into adhesive contact with the polymer surface. Furthermore, we showcase the versatility of the approach by placing two nanorods on top of each other and by depositing high aspect ratio InAs nanowires across electrodes.

**Results and Discussion.** The assembly process is shown in Figure 1. Thermal scanning probe lithography (t-SPL) is

Received: September 6, 2019
Revised: October 30, 2019
Published: November 6, 2019ACS Publications   © 2019 American Chemical Society   8855   DOI: 10.1021/acs.nanolett.9b03687
Nano Lett. 2019, 19, 8855−8861This is an open access article published under an ACS AuthorChoice License, which permits copying and redistribution of the article or any adaptations for non-commercial purposes.

Downloaded via IBM CORP on September 10, 2020 at 14:41:10 (UTC).
See https://pubs.acs.org/sharingguidelines for options on how to legitimately share published articles.

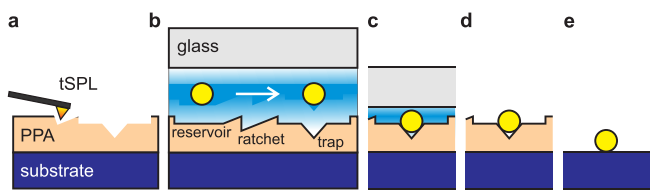

**Figure 1.** Process flow of nanoparticle assembly by nanofluidic confinement. (a) A topography is patterned into a PPA film by thermal scanning probe lithography (tSPL). (b) Particles are transported by a rocked Brownian ratchet (white arrow) from a reservoir to assembly sites, where they diffuse to individual traps in the PPA film. (c) The gap distance is reduced until the particles jump into adhesive contact with the trap. (d) The suspension is removed by rinsing, and the van der Waals force keeps the particle at the desired location. (e) The PPA film is removed by self-amplified depolymerization at ∼220 °C on a hot plate.

used to pattern a PPA film, which serves as a removable template[10] (Figure 1a). The pattern gives rise to an energy landscape for the particles in a nanofluidic slit. The landscape enables transport[18] of the particles from a reservoir to assembly sites and trapping and orientation of the particles at the target positions (Figure 1b). Finally, we bring the particles into adhesive contact with the polymer for immobilization (Figure 1c). After drying, the template is thermally removed (Figure 1d,e). We note that this step proceeds without a detectable shift in the lateral particle position.[10]

*Accurate Deposition of 60 nm Au Spheres.* The pattern used for trapping and depositing 35 Au spheres of 60 nm diameter is shown in Figure 2a. It consists of 35 nanoscale indents within a rectangular compartment. The compartment and the holes are ∼22 and 10 nm deep, respectively; see Figure 2b. The lower part of Figure 2a also depicts the exit channel of the Brownian motor ratchet device which was used to load the spheres into the compartment. The Brownian motor transport scheme was described in detail in a previous publication,[18] and a short description is given in the Supporting Information. Here we focus on the trapping and immobilization steps.

After depositing a drop of the nanoparticle suspension on the sample, a cover glass is brought into close proximity with the polymer surface using the previously described nanofluidic confinement apparatus.[21,22] Using citrate stabilized 60 nm diameter gold spheres from BBI solutions (HD.GC60.OD100, diluted 1:20 with Millipore water, 8% polydispersity), we found that all surfaces are negatively charged and the system is stable. In the experiments, the average gap distance for all recorded images can be detected and measured with high precision from the interference of light reflected from the glass–water interface and the sample surface as described earlier.[21,22] At a gap distance of $d \approx 120.2 \pm 0.5$ nm (the error accounts for the standard deviation of the gap distances in the recorded movies) between the unpatterned polymer surface and the cover glass, the energy barriers are small and the particles explore the entire area of the compartment by diffusion. Figure 2c depicts the average particle occupation statistics of five particles recorded over a period of 13 s (13,000 movie frames) in the form of a heat map.

Clear hot spots of the particle density could be identified, as depicted by the blue circles in Figure 2c. We omitted the right side of the compartment for this analysis, since in this area the particle density was too high and faithful particle tracking was not possible. The average hot spot image is shown in Figure 2d with a pixel resolution of 5 nm. The image was obtained by first determining the center of each hot spot by radial-symmetry-based tracking[23] and collecting all recorded particle positions relative to the hot spot center. For comparison, Figure 2e shows the average trap topography that was obtained by cross-correlation averaging. Both images reveal that the traps were slightly elongated in the x-direction, likely caused by an asymmetric writing process. The asymmetry is small, and therefore, we used the particle distribution to calculate an empirical radial probability density $P(r)$; see the blue dots in Figure 2f. The black circles depict the averaged radial topography profile. The width of the Gaussian fit (red line) to the topography data is $\sigma_z = 17.3 \pm 0.5$ nm. We note that, due to the finite size of the imaging tip, this value is a lower bound.

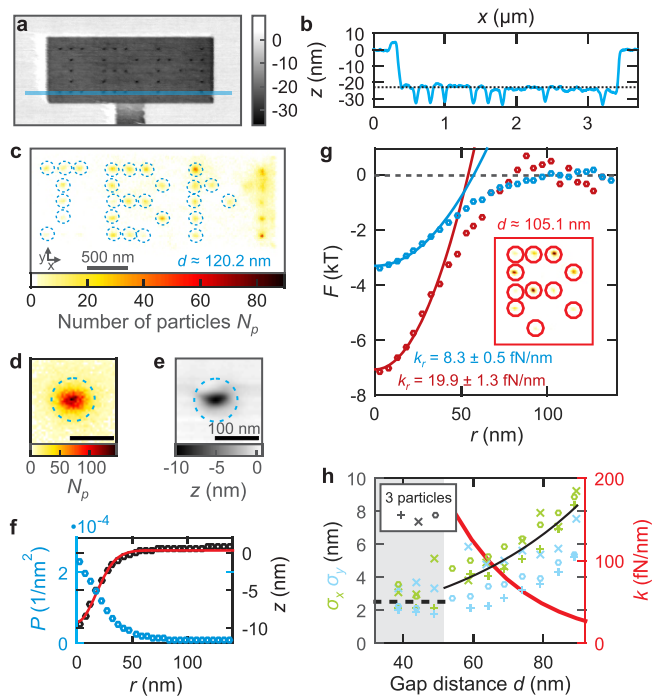

**Figure 2.** Geometry-induced electrostatic trapping. (a) Topography image of the assembly area. The opening at the bottom is the entrance from the Brownian motor. (b) Cross section measured across the blue line in panel a. The area is recessed by ∼23 nm and contains traps of ∼10 nm depth. (c) Heat map (10 nm grid) of detected particles $N_p$ from ∼13,000 frames. (d) The average particle distribution from locations marked by the blue circles in part c (5 nm grid). (e) Average topography of the traps in part a. (f) Average empirical radial occupation probability (blue) and topography (black). The red line is a Gaussian fit ($\sigma_z \approx 17$ nm). (g) Ensemble averaged free energies $F(r)$ at gap distances of $d \approx 120.2$ nm (blue) and $d \approx 105.1$ nm (red). The inset shows the particle distribution for $d \approx 105.1$ nm. The solid lines indicate parabolic fits. (h) Standard deviations of the measured positions of three spheres in the x- and y-direction, marked by green and blue symbols, respectively. The extrapolated trap stiffness is depicted by the red line. The black line marks the calculated $\sigma_{x,y}$ inferred from the trap stiffness. In the gray region, the particles are immobilized. In this region, an averaged lateral localization accuracy of $\sigma_{x,\text{noise}} \approx 2.5$ nm (dashed line) is determined.

Next, we derive the free energy $F(r)$ according to the Boltzmann relation by $F(r)/k_B T = -\ln(P(r))$, where $k_B$ is the Boltzmann constant and $T$ is the absolute temperature. The free energy $F$ is shown in Figure 2g by the blue and red circles for a confinement of $d \approx 120.2$ nm and $d \approx 105.1$ nm, respectively. In first order, the trap acts like an elastic spring which pulls the particles back into the center. The spring stiffness also known as the trapping stiffness was assessed by fitting the parabolic energy profile of a harmonic oscillator to the free energy curves for $r < 40$ nm. The trapping stiffnesses are $k_r = 8.3 \pm 0.5$ fN/nm and $k_r = 19.9 \pm 1.3$ fN/nm, as depicted by the blue and red lines, respectively.

We note that the measurements of the trapping energy and stiffness are affected by the finite particle illumination time of ~50 μs, which leads to an averaging of their particle position.[24] We describe in the Supporting Information how the values for trapping stiffness were corrected for the finite illumination time and laser noise. The corrected values for the trapping energy $F(0)$, the standard deviation of particle positions $\sigma_r$, the trapping stiffness $k_r$, and the relaxation time $\tau_{rel}$ are shown for both gap distances in Table 1.

Table 1. Values for the Trapping Energy $F(0)$, the Standard Deviation of Particle Positions $\sigma_r$, the Trapping Stiffness $k_r$, and the Relaxation Time $\tau_{rel}$, Corrected for the Finite Illumination Time and Laser Noise

| $d$ | 120.2 nm | 105.1 nm |
| --- | --- | --- |
| $F(0)$ | $-3.1 \pm 0.2\ k_BT$ | $-6.7 \pm 0.4\ k_BT$ |
| $\sigma_r$ | $24.3 \pm 1.4$ nm | $17.0 \pm 1.1$ nm |
| $k_r$ | $6.9 \pm 0.4$ fN/nm | $14.2 \pm 0.9$ fN/nm |
| $\tau_{rel}$ | $81 \pm 5$ μs | $40 \pm 3$ μs |

In previous experiments, an exponential increase of the potential energy[18] and the trapping stiffness[14] with decreasing gap distance was observed. Assuming an exponential decay, i.e., $k_r = A\exp(-Bd)$, we obtain parameters of $A \approx 1.96$ pN/nm and $B \approx 0.048$ 1/nm by using the corrected trapping stiffnesses at a gap distance of $d \approx 105.1$ nm and $d \approx 120.2$ nm (see the red line in Figure 2h). Using the equipartition theorem, we can now determine the expected variance $\sigma_r^2$ of the particle motion in the traps according to[24]

$$\sigma_r^2 = \frac{k_BT}{k_r} \tag{1}$$

We measured the standard deviations of the particle position for three particles in the $x$- and $y$-direction as a function of the gap distance; see the green and blue symbols in Figure 2h. Good agreement with the expected values (black line) is found.

We also observe a systematically increased standard deviation in the $x$-direction compared to the $y$-direction, consistent with the asymmetric traps. Clearly, the increasing trapping stiffness leads to a focusing effect and thus to a more precise deposition process.

*Details of the Immobilization Process.* In iSCAT imaging, the observed contrast of a particle depends on its vertical position in the nanofluidic gap and can be used to measure its vertical position.[21,22] Figure 3a shows the normalized contrast of a single particle for varying gap distances $d$. At a gap distance of $d \approx 90$ nm, the particle diffuses into a trap, as schematically depicted in Figure 3a(i). Upon closing of the gap, the contrast value distribution becomes first more narrow until the contrast changes abruptly to a higher value. We interpret this sudden change as a jump into contact with the polymer surface; see Figure 3a(ii). Indeed, after reaching this point, the contrast fluctuations have a similar width, as measured for immobilized particles. At a gap distance of $d \approx 35$ nm, the motion of the cover glass is reversed, and for $d > 40$ nm, a hysteresis in the measured contrast is observed. In this region (see blue box in Figure 3a), the contrast change is steep. We interpret this behavior as follows: At $d = 40$ nm, the particle touches both surfaces. Upon further decreasing the gap distance, the particle is pressed into the polymer and the system deforms mostly elastically. Upon retracting the glass, first this elastic deformation is restored and the contrast is reversible. Then, the contrast differs between the two directions because the particle is now fixed on the polymer surface. Upon further retraction of the glass surface, the observed contrast modulation (yellow region) is small, similar to the case of particles immobilized on a silicon sample.[21]

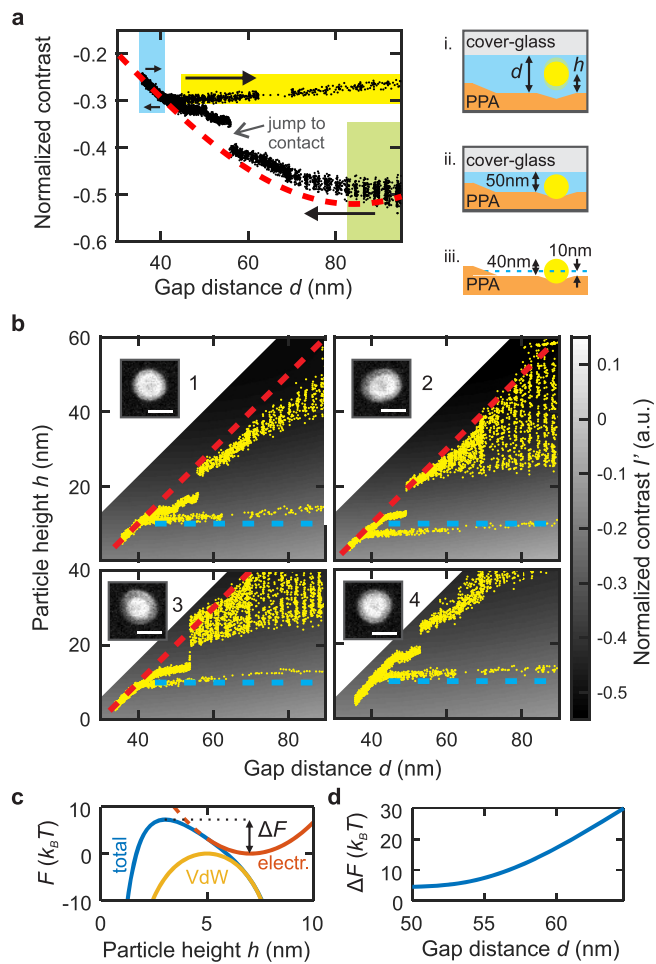

Figure 3. Particle deposition. (a) Normalized contrast (black scatter plot) of a single trapped particle for varying gap distances. The black arrows indicate if the cover glass approaches the PPA layer or if it is retracted. The dashed red line indicates the simulated minimal contrast. At a gap distance of $d \approx 53$ nm, the contrast makes a jump and then stays almost constant, while the cover glass position is lowered. The contrast signal has a hysteresis after the particle is in contact with both surfaces (blue region), and the contrast is only slightly modulated while retracting the cover glass (yellow region). The right insets illustrate the measured particle height $h$ for several gap distances $d$. (b) Attributed particle heights (yellow dots) are obtained by matching the measured contrast and simulated contrast values (gray scale background image). The four panels show the attributed heights for four different particles. (c) DLVO model curves for a gap distance of 55 nm. $\Delta F$ denotes the energy barrier assuming that hydration layers stabilize the particle–glass interface. (d) $\Delta F$ as a function of the gap distance.

Using this interpretation, we can now adapt the optical model for the particle contrast described elsewhere[21] to this system. For this model, we fit the minimum contrast predicted for all possible particle heights (dashed red line) to the observed data using the three free parameters of the model. In the blue region, the particle is in contact with the glass surface. Moreover, in the green region, the particle contrast should have a minimum. After optimization, we obtain the three fit parameters for the four particles in Figure 3b; see Table 2. With these parameters, we convert the contrast data to the particle height data; see Figure 3b.

Table 2. Optical Fit Parameters of the Four Particles in Figure 3b, Where $\gamma$ Is the Fraction of the Incoming Light Interacting with the Particle, $p$ Is the Scattering Amplitude of the Particle, and $\phi_0$ Defines the Effective Scattering Phase

|  | 1 | 2 | 3 | 4 |
| --- | --- | --- | --- | --- |
| $\phi_0$ ($\pi$) | 0.86 | 0.87 | 0.84 | 0.86 |
| $\gamma$ | 0.13 | 0.12 | 0.11 | 0.12 |
| $p$ | 0.74 | 0.95 | 0.82 | 0.66 |

The insets in Figure 3b depict the respective SEM images of the deposited particles. Particles 2 and 3 are nonspherical, which results in an asymmetric scattering cross section and explains the wide scatter of apparent particle heights at $d_c \lesssim d$ due to the rotation of the particles. For particle 2, the asymmetry of the traps prevented the particle from rotating in the $xy$-plane at smaller gap distances and it aligns along the wider $x$-direction of the traps, while particle 3 still rotates just before the jump to the polymer surface.

All particles show a jump into contact with a jump distance of 5−9 nm at an average gap distance of $d$ = 49−56 nm. The size of this jump is consistent with a simple Derjaguin, Landau, Verwey, Overbeek (DLVO) model of a particle between two planar surfaces with a distance of $d$ + 15 nm; see Figure 3c calculated for $d$ = 55 nm. The model assumes a Debye length of 12.5 nm, as obtained from conductivity measurements, and a surface potential of −67 mV for the polymer[22] and −45 mV for the glass. The potential of the particle was assumed to be −58 mV, as measured for the zeta potential of the particles. Moreover, a Hamaker constant of 5.7 × 10$^{-20}$ J was assumed for the silica−water−gold system[25] and used for both interfaces. The reduced potential of −45 mV of the glass interface (from −67 mV[26]) is required in order to explain the assymmetric presence of the particles close to the glass surface. This asymmetry was observed before in the same system.[22] Moreover, it is likely that charge regulation processes at the glass interface are effective due to the close proximity of the particle of less than 1 Debye length.[26] Furthermore, according to the DLVO model, the size of the van der Waals energy is sufficient to pin the particles to the glass surface. However, additional hydration layer forces may stabilize the particle−glass interface[27] and could explain the observed jump away from the glass surface. Assuming a stabilized glass interface, the model correctly predicts barriers $\Delta F$ of less than $10k_BT$ at a gap distance of less than 57 nm, consistent with the observed jump positions.

After the jump into contact, the particle's center position is 15−19 nm above the surface, which corresponds to a penetration of 11−15 nm into the polymer, consistent within experimental error with the trap depth of 10 ± 2 nm and a conformal adaptation of the polymer−particle interface upon contact.[28] In addition to the absence of hydration layers at the more hydrophobic polymer surface (contact angle ≈ 63°), such a conformal contact would also explain why the particles jump only once to the polymer surface and do not oscillate between the interfaces, as would be expected from the similar Hamaker constants of both materials.

At a sufficiently large gap distance, other particles started to fill the device area by diffusion and the transport, trapping, and assembly process was repeated. With each approach, we assembled approximately five to eight particles. This number was limited by the amount of particles delivered to the assembly site and can be adapted by increasing the particle concentration in the suspension.

Figure 4a depicts an AFM image of the particles after deposition. Consistent with the discussion above, the particles protrude 40 nm from the surface. We also imaged the particles by SEM; see Figure 4b. The 34 red plus signs indicate the central position of correctly placed nano-objects, as determined by radial-symmetry-based tracking.[23] The black plus sign shows the position of one particle which was deposited at a wrong location, while one trap in the letter I is vacant. Also, an agglomerate of two particles was assembled in the letter M. The target positions of the traps were located on a 200 nm grid. Please note that SEM imaging is performed by raster scanning, which is prone to thermal drift and charging. To compensate this effect, we performed a least-squares fit of a skewed and $y$-scaled 200 nm squared grid to the data of detected particle positions. The obtained trap positions are shown by the blue circles in Figure 4b.

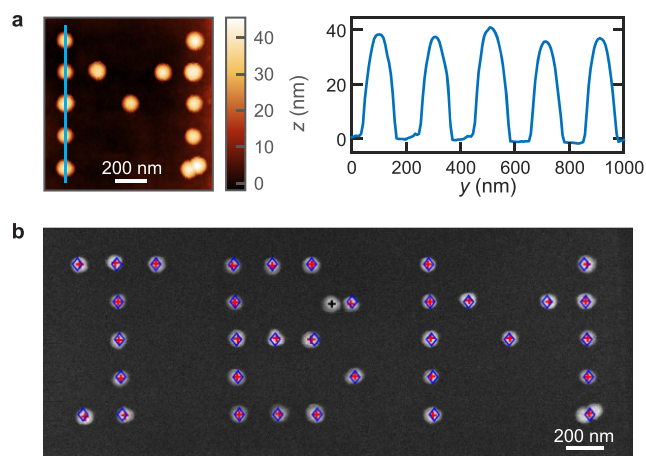

Figure 4. Assembled 60 nm Au spheres. (a) Topographical AFM scan of particles assembled to form the letter M. The cross-sectional profile along the blue line indicates that particles stuck out 40 nm from the bottom surface of the assembly site. (b) SEM image of all assembled particles. The centers of correctly assembled particles and their grid position are shown by red plus signs and blue diamonds, respectively. A black plus sign indicates the position of a particle at a wrong location.

As a measure of positioning accuracy, we determined the distances $r$ between the centers of the nanospheres and the traps. The standard deviation of the displacement is $\sigma_x$ = 6.6 nm and $\sigma_y$ = 5.2 nm in the $x$- and $y$-directions ($\sigma_r$ = 8.3 nm), consistent with the asymmetric traps. The values are slightly larger than the observed standard deviations during assembly, which may be caused by nonlinear scanning artifacts during imaging or position jitter during patterning.

*Stacked Assembly and Assembly of High Aspect Structures.* In the previous section, we used spheres to elucidate how particles become immobilized when the gap distance is decreased. In addition, the shape of the traps can be matched to the shape of nanoparticles to be assembled. Moreover, t-SPL can be used to image pre-existing structures on the surface and to position the trap in registry with those structures. Here we present two showcase examples which exploit these functionalities.

In plasmonic applications, it is desirable to position metallic features in close proximity and in three dimensions in order to exploit the coupling of the metal structures, for example, for the fabrication of meta materials. Here we demonstrate that two silver nanoparticles can be stacked on top of each other. We used pentagonal silver nanorods with a plasmon peak at ∼780 nm and a diameter of ∼45 nm (Sciventions).

After immobilizing nanorods on a silicon wafer by droplet drying, a 65 nm thick PPA film was spin-coated on top of the sample. Similar to the case of a buried nanowire,[29] a residual topography on the spin-coated film marks the precise location of the buried nanorod (see Figure 5a). Next, a trap was written with the long axis perpendicular to the buried rod by t-SPL; see Figure 5b. Using this topographical trap and the assembly process described above, a second particle was placed on top of the first assembled particle (see Figure 5c). After drying, the sample was inspected by SEM.

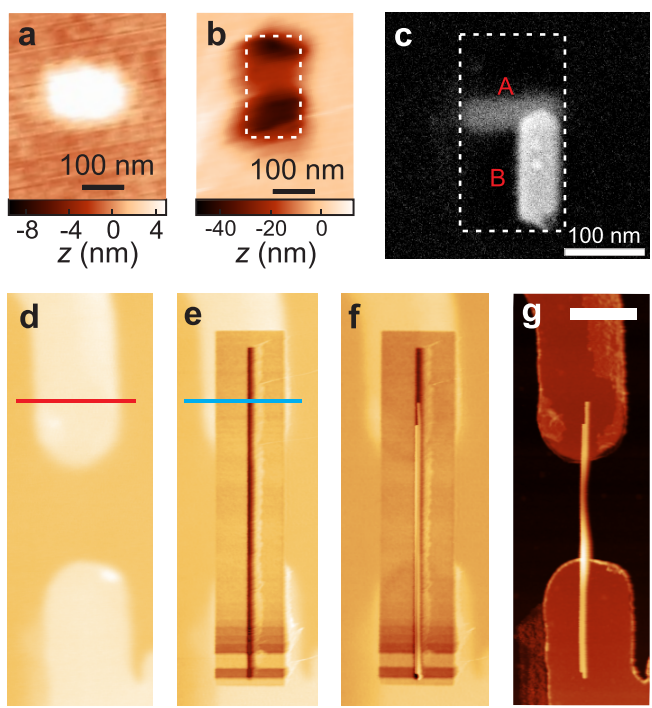

**Figure 5.** Application range demonstration. (a) t-SPL image of a gold nanorod buried in a 60 nm thick PPA film. (b) Trap definition on top of the nanowire by t-SPL. (c) SEM image of the final assembly of two nanowires on top of each other. (d) t-SPL image of ∼25 nm thick electrodes covered by a 135 nm thick PPA film. (e) Same area after patterning the nanowire traps, (f) after deposition of a nanowire bundle, and (g) after removal of the polymer.

The second rod aligned well to one sidewall of the trap at a position of maximum trap depth. The overall misalignment of ∼4° from the perpendicular direction is due to a slight misalignment of the trap with respect to the first rod (see Figure 5c). The lateral displacement was around 80 nm, which could easily be improved by a narrower and shallower trap geometry. Also, the distance between the rods could be adjusted by a precise measurement of the polymer thickness and an accurate depth control when placing the second trap.

As a second showcase, we demonstrate the placement of high aspect ratio nanowires on predefined electrodes. These single-crystalline wires offer ballistic transport over more than 100 nm.[30] A precise placement on predefined electrodes on a substrate is a first step toward assembled high performance electronic circuits. We used a highly doped silicon substrate with a 135 nm thick layer of thermal $SiO_2$ and 5 nm of evaporated $HfO_2$. Subsequently, the electrodes were defined by optical lithography and a lift-off process using an ∼5 nm Cr adhesion layer and an ∼20 nm thick evaporated gold layer. The metal layers were patterned into several pairs of ∼1 μm wide electrodes, which face each other at a distance of ∼1 μm. In a next step, a 135 nm thick PPA film was spin-coated on top of the substrate.

The surface topography imaged with the t-SPL tip is shown in Figure 5d. Figure 5e shows the topography after the t-SPL patterning. The elongated traps were ∼5 μm long and had a maximum depth of ∼40 nm.

The InAs nanowires were grown epitaxially on a (111) InAs substrate by using gold seed particles, which act as collectors of vapor-phase precursor materials.[31] The growth was initiated as soon as the precursor gases arsine ($AsH_3$) and trimethylindium ($TMIn$) were available at a growth temperature of 430 °C. The wires were coated with 3 nm $Al_2O_3$ and 3 nm $SiO_2$ using atomic layer deposition (ALD). The high-κ dielectric $Al_2O_3$ is well suited for gating, yet it has an isoelectric point[32] close to pH ≈ 9. Therefore, to obtain a negative charge on the wires, we added a silicon oxide layer, which has an isoelectric point[32] around pH ≈ 2.

The wires had a density of ≤9 per μm² on the wafer, a length of ∼4.3 μm, and a diameter of ∼35 nm. The wires were harvested by first freezing a drop of ultrapure water (Millipore) on a small substrate piece. The drop was frozen by letting the sample float in a metal dish on liquid nitrogen. The frozen drop was then sheared off with tweezers and used for assembly.

The nanowire suspension was placed on the patterned sample and confined in the NCA. Naturally, diffusion for such large objects is rather slow and we used electo-osmotic forces to control the position of the wires and to drag them toward the trap; see the Supporting Information for details. Once the nanowire was close to a trap, we reduced the gap distance from $d ≈ 380$ nm to $d ≈ 300$ nm, which ensured a stable trapping of the wires. Figure 5f shows the assembled nanowire bundle inside the trap. In subsequent transport and assembly steps, two more single nanowires and two more bundles were placed in similar traps (see the Supporting Information).

All wires were transferred onto the electrodes by placing the sample for 10 s on a hot plate with a temperature of ∼220 °C. This temperature was found to reliably decompose the PPA film without residues. Figure 5g shows an AFM topography image of the assembled nanowires.

**Discussion and Outlook.** We demonstrated that geometry-induced trapping enables the trapping and deposition of nanoparticles and high aspect ratio nano-objects with correct orientation at a target location. Upon a controlled decrease of the gap distance, the traps become stiffer, which leads to a focusing of the particles just before they jump into contact with the template polymer surface. This mechanism explains the high accuracy of the method of approximately 5 nm in the x- and y-direction. Moreover, we demonstrate stacked assemblies—interesting for plasmonic applications—and deposition of large aspect ratio nanowires with high accuracy.

In contrast to dielectrophoretic deposition, functional electrodes are not required and a high density and/or a complex arrangement of single nanowires into functional electronic circuits is within reach, a first step toward future nanoprocessors.[2] For such a goal, however, the stability of the wires in solution must be enhanced, for example, by using shorter wires and basic pH conditions in order to increase the negative charge on the wires. Moreover, in our experiments, the transport of the wires to the traps is time-consuming and more efficient transport schemes are required to increase the throughput. A possibility is to use fluid flow for a fast transport of the nano-objects toward the trap arrays. It has been shown for gold nanospheres that a high yield filling of electrostatic traps can be achieved.[12] Such a combination could lead to a larger scale parallel deposition method for high aspect ratio nano-objects into functional electronic circuits.

## ■ ASSOCIATED CONTENT

### *ⓢ* Supporting Information

The Supporting Information is available free of charge on the ACS Publications website at DOI: 10.1021/acs.nanolett.9b03687.

> Details related to overall device design and Brownian transport to the compartment, correction of motion blur and noise, and nanowire harvesting and transport to the target sites (PDF)

## ■ AUTHOR INFORMATION

**Corresponding Author**
*E-mail: ark@zurich.ibm.com Phone: +41 (0)44 7248246.
**ORCID**
Armin W. Knoll: 0000-0003-2301-3149

**Notes**
The authors declare no competing financial interest.

## ■ ACKNOWLEDGMENTS

The authors thank U. Drechsler for assistance in fabrication, H. Schmid for the growth of the InAs nanowires, and R. Allenspach and H. Riel for support. Funding was provided by the European Research Council (StG No. 307079 and PoC Grant No. 825794) and the Swiss National Science Foundation (SNSF No. 200020-144464 and No. 200021-179148).

## ■ REFERENCES